\documentclass[journal]{IEEEtran}
\ifCLASSINFOpdf
\else
\fi

\usepackage{array}
\usepackage{graphicx}
\usepackage[square,comma,sort&compress,numbers]{natbib}
\usepackage{amsmath}
\usepackage{amssymb}
\graphicspath{{Figures/}}
\begin{document}
\title{Massive Non-Orthogonal Multiple Access for Cellular IoT: Potentials and Limitations}
\author{Mahyar~Shirvanimoghaddam,~\IEEEmembership{Member,~IEEE,}
                 Mischa~Dohler,~\IEEEmembership{Fellow,~IEEE,}
                Sarah~J.~Johnson,~\IEEEmembership{Member,~IEEE}

\thanks{M. Shirvanimoghaddam and Sarah. J. Johnson are with School of Electrical Engineering and Computer Science, The University of Newcastle, NSW, Australia.

M. Dohler is with King's College London, UK.}}
\maketitle

\begin{abstract}
The Internet of Things (IoT) promises ubiquitous connectivity of everything everywhere, which represents the biggest technology trend in the years to come. It is expected that by 2020 over 25 billion devices will be connected to cellular networks; far beyond the number of devices in current wireless networks.  Machine-to-Machine (M2M) communications aims at providing the communication infrastructure for enabling IoT by facilitating the billions of multi-role devices to communicate with each other and with the underlying data transport infrastructure without, or with little, human intervention. Providing this infrastructure will require a dramatic shift from the current protocols mostly designed for human-to-human (H2H) applications. This article reviews recent 3GPP solutions for enabling massive cellular IoT and investigates the random access strategies for M2M communications, which shows that cellular networks must evolve to handle the new ways in which devices will connect and communicate with the system. A massive non-orthogonal multiple access (NOMA) technique is then presented as a promising solution to support a massive number of IoT devices in cellular networks, where we also identify its practical challenges and future research directions.
\end{abstract}

\IEEEpeerreviewmaketitle

\section{Introduction}
\IEEEPARstart{I}{nternet} of Things (IoT) is one of the biggest technology trends which aims at transforming every physical object to an information source. IoT use cases can be generally divided into two large categories. In Massive IoT applications, sensors typically report to the cloud on a regular basis, and the requirement is for low-cost devices with low energy consumption and good coverage. Examples include smart buildings, logistics, tracking, and fleet management. In critical IoT use cases, there are high demands for reliability, availability and low latency. Critical IoT includes remote health care, traffic safety and control, industrial applications and control, remote manufacturing, training and surgery. The continued fall in the price, size, and power consumption, of autonomous devices capable of sensing and actuating, have been the driving force for increasing the popularity of IoT systems and services.

Ericsson forecasted that the IoT will include over 25 billion units installed by 2020 and a large share of these will be applications serviced by short-range radio technologies, such as WiFi, Bluetooth, and Zigbee with limited quality of service (QoS) and security requirements typically applicable for indoor environments, while a significant proportion will be enabled by wide area networks mostly facilitated by cellular networks \cite{M2M_Ericsson2}. By 2020, IoT product and service suppliers will also generate incremental revenue exceeding \$300 billion, mostly in services \cite{M2M_Ericsson2}. As operators are responsible for wireless connectivity on a global scale, they are in an excellent position to participate in the IoT market and capture a share of the added value generated by the emerging IoT applications.

\begin{figure}[t]
\centering
\includegraphics[width=\columnwidth]{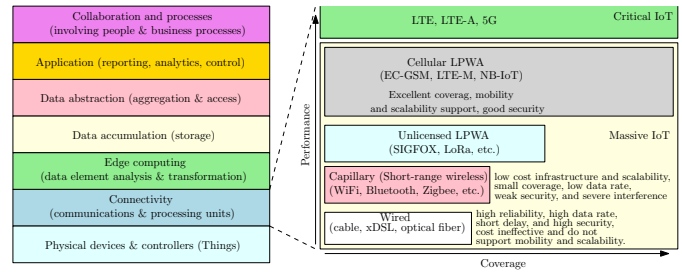}
\caption{IoT reference model.}
\label{fig:IoTRef}
\end{figure}

Machine-to-machine (M2M) communications refer to automated data communications among machine type communication (MTC) devices and constitutes the basic communication paradigm in the emerging IoT \cite{TUbiq}. The IoT reference model \cite{M2M_America} is shown in Fig. \ref{fig:IoTRef}, which shows that connectivity is the essential part of the entire IoT ecosystem, which can be provided through wired or wireless solutions.  Connectivity through cellular networks is facilitated through the third generation partnership project (3GPP) technologies, including GSM, WCDMA, LTE and future 5G. These technologies operate on licensed spectrum and are primarily designed for high quality mobile voice and data services. They are now being evolved with new functionality to form attractive solutions for emerging low-power IoT application. 3GPP technologies already dominates many IoT applications which require large geographical coverage and medium-to-high performance requirements \cite{M2M_Ericsson2}. The key challenges for massive IoT deployment in cellular networks include, \emph{1) device cost}, which is the enabler for high volume mass-market applications, \emph{2) battery life}, to reduce the cost of replacing batteries, \emph{3) coverage}, deep indoor connectivity and and regional coverage is a requirement for many IoT applications, \emph{4) scalability}, the network capacity must be easily scaled to handle millions of devices, and \emph{5) diversity}, connectivity should support diverse range of service requirements; e.g., alarm signals applications require highly reliable communication with QoS guarantees, while in smart metering applications delays are typically tolerable \cite{M2M_America}.

3GPP has made significant improvements to meet the requirements of emerging massive IoT applications, which leads to a range of cellular low-power wide area solutions. These new solutions include extended coverage GSM (EC-GSM) which is achieved by defining new control and data channels mapped over legacy GSM, narrow band IoT (NB-IoT) which is a self-contained carrier that can be deployed with a system bandwidth of 200 kHz and is enabled on an existing LTE network, and LTE for MTC (LTE-M) which brings new power-saving functionality to LTE suitable for M2M applications. The most important improvements which 3GPP has made so far to enable massive IoT are 1) lower device cost by reducing peak rate, memory requirements, and device complexity, 2) improved battery life up to 10 years, by introducing power saving mode and discontinues reception, and 3) improved coverage, e.g., 15 dB and 20 dB in link budget on LTE-M and NB-IoT, respectively \cite{M2M_Nokia} which is equivalent to the signal penetrating a wall or floor, enabling deeper indoor coverage. Despite huge efforts of 3GPP towards making M2M communications a reality, there are still open challenges where efficient solutions must be proposed.

As part of massive IoT supports, a very large number of MTC devices, few million devices per square kilometer, must be supported. Towards this, non-orthogonal multiple access (NOMA) has been identified as a key technique in 5G which can enable trillions of MTC devices to communicate with the base station in cellular IoT use cases \cite{M2M_America}.  Also mentioned in \cite{M2M_Nokia}, grant-free uplink through resource spread multiple access enables asynchronous, non-orthogonal contention-bases access that is well suited for sporadic uplink transmissions of small data bursts common in IoT use cases which is considered as new capabilities for the massive IoT in 5G. The presented NOMA strategy in this paper shows how NOMA can be used on top of existing cellular infrastructure to enable massive number of devices to share the same radio resources; thus enabling massive IoT applications.

\section{Current Access Techniques}
\begin{figure}[t]
\centering
\includegraphics[width=0.5\columnwidth]{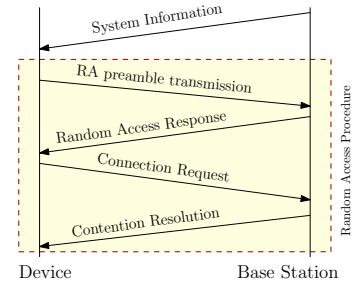}
\caption{Random access procedure in LTE.}
\label{fig:RALTE}
\end{figure}

In most existing wireless networks, radio resources, including time and frequency, are \emph{orthogonally} allocated to different devices for data transmission. The process of devices contacting the base station to request a   transmission slot is called the random access (RA) procedure. In the current LTE standard, RA is a four step handshake process which is depicted in Fig. \ref{fig:RALTE}.
\subsection{The Random Access Procedure}
The devices are first informed of the available physical random access channel (PRACH) resources, comprising of a periodic amount of time-frequency resources, through the system information broadcasted by the base station. In the first step of the RA procedure, each device randomly chooses a preamble among available set of 64 preambles and sends it to the base station. The base station can then detect the transmitted preambles by calculating the cyclic cross correlation of the set of preambles with the received signal and using the result can estimate the transmission time of devices which have selected each preamble. Upon detecting each preamble, the base station sends a random access response (RAR) message, including the information of the radio resources allocated to devices and the timing advance information for all the devices which have selected a specific preamble, to adjust synchronization. Once a device has received the RAR message, it sends a temporary terminal identity through the allocated radio resource to the base station to request a connection. The base station sends information allocating radio resources to each of the devices that have gained access by specifying their terminal identity. Therefore, a connection is established and the device can start the data transmission in allocated time-frequency slots.

\subsection{Challenges of Conventional Random Access for Massive Cellular IoT}
The random access procedure in current cellular standards is only feasible when the number of devices is small enough that\emph{ i)} the devices don't unduly interfere with each other in the random access phase and \emph{ii)} there is sufficient radio resources to allocate a separate data channel to each user in the transmission phase. In the following we outline the most significant challenges of the conventional RA procedure for massive cellular IoT.

\textbf{\emph{Preamble collision and overload problems:}}  Two conditions may arise in the random access procedure which dramatically limit its efficiency: (1) The condition known as \emph{preamble collision}, where more than one device selects the same preamble in the first step of the RA procedure, causing co-channel interference in the transmission stage and (2) the condition known as \emph{overload} which is due to the SINR violation caused by an excessive number of transmissions by other nodes in the same cell or in neighboring cells during the RA procedure. In case of collision, the devices will repeat the preamble transmission in the next available RA resource. Frequent preamble retransmissions however leads to network congestion, increasing delays, packet loss, high energy consumption, excessive signaling overhead, and radio resource wastage \cite{M2MRA}.  Existing solutions to solve the collision problem use different approached to delay the retransmission of preambles in order to minimize the collision probability. These include \textit{dynamic allocation} \cite{MTCSurvey}, \textit{slotted access}, \textit{group-based} \cite{TUbiq}, \textit{pull-based}, and \textit{access class barring} \cite{ClassBarr}. Although these approaches can reduce the access collision to a certain degree, they are unable to support a large number of devices in IoT scenarios \cite{ChallM2MAccess}.

\textbf{\emph{Excessive overhead:}} Another problem with the connection-oriented communication in the current LTE standard is \emph{\textbf{excessive signalling overhead}} as significant resources are spent establishing a connection to allow transmission of the very small-sized data (e.g., a few kbs) typically required for M2M communications, especially when a large number of M2M devices attempt to access cellular networks at the same time \cite{HybridRAandDataM2M}. For example, to transmit 100 bytes of data, approximately 59 bytes of overhead on the uplink and 136 bytes on the downlink would typically be required for signalling transmissions \cite{MTCSurvey}. Hybrid schemes are proposed to combine the RA procedure and the data transmission, where the devices will send their messages through the third message of the RA procedure. Data aggregation could also be used for more efficient transmission which in only applicable for delay tolerant M2M applications. These strategies are also not scalable and mostly inefficient when the number of device is very large.

\textbf{\emph{Different QoS requirements:}} Many M2M applications have diverse service requirements which must be carefully considered when designing the access techniques. For example, some M2M applications, such as alarm signals, are delay sensitive and a very small message must be delivered within 10 msec, while other applications, such as smart metering, are delay tolerant or have larger packet sized and can tolerate delays of several hours.  Most existing access techniques for M2M communications have not considered QoS requirements of MTC devices and treat all the devices as the same. This is however ineffective for real world M2M applications and leads to huge radio resource wastage and/or service interruptions. QoS can be effectively taken into account when designing the access technology especially in massive IoT applications, which can significantly reduce the load on the core network and minimize the access delay for delay sensitive applications.

\textbf{\emph{Co-existence with H2H devices:}} In cellular based M2M applications, MTC devices coexist with H2H devices. Since the number of MTC devices is very large compared to that of H2H and they are using the same radio resources, inefficient design of the transmission procedure for MTC devices may result in huge losses in H2H communication. One could define new PRACH resources for M2M devices to avoid congestion with H2H, or consider dynamic PRACH resource allocation to adjust available resources based on the estimated traffic. Although these approaches can maintain the QoS requirements for H2H devices, many M2M devices may be required to delay their transmissions and wait until an appropriate number of resources become available for M2M devices. This problem will be more challenging when the number of M2M devices is very large and only a limitted number of radio resources is available for both H2H and M2M traffic.
\section{Non-orthogonal Multiple Access for M2M Communications}
\begin{figure}[t]
\centering
\includegraphics[width=\columnwidth]{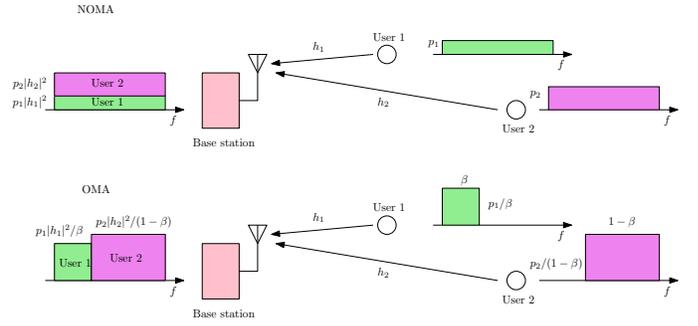}
{\footnotesize (a) NOMA vs. OMA}
\includegraphics[width=\columnwidth]{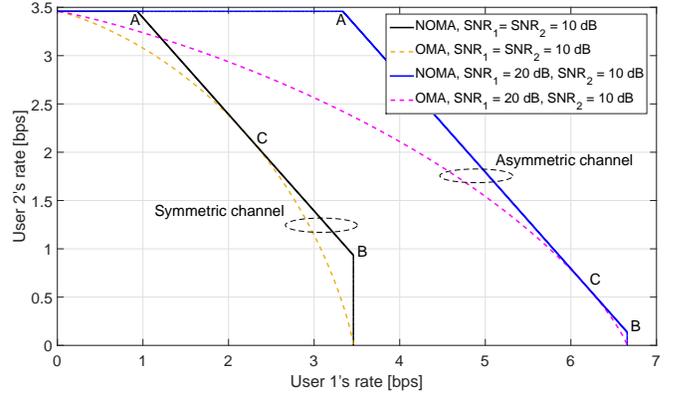}
{\footnotesize (b) Capacity region of the multiple access channel.}
\caption{NOMA and OMA and their achievable rate for a 2 user scenario.}
\label{fig:NOMA_OMA}
\end{figure}
Multiple access techniques can be classified into orthogonal and non-orthogonal approaches. In orthogonal multiple access (OMA), including time division multiple access (TDMA), frequency division multiple access (FDMA), and orthogonal FDMA (OFDMA), signals from different users are not overlapped with each other. Non-orthogonal schemes however allow overlapping among the signals in time or frequency by exploiting power domain, code domain or interleaver pattern, often providing better performance in comparison with orthogonal schemes in terms of throughput \cite{China_NOMA}. Orthogonal multiple access is a suitable choice for packet domain services with channel aware time and frequency scheduling \cite{China_NOMA}. However, further improvements in the system efficiency and QoS required for the fifth generation (5G) of mobile cellular networks and IoT applications, necessitates the adoption of NOMA schemes with high throughput efficiency.
\begin{figure*}[t]
\centering
\includegraphics[width=1.7\columnwidth]{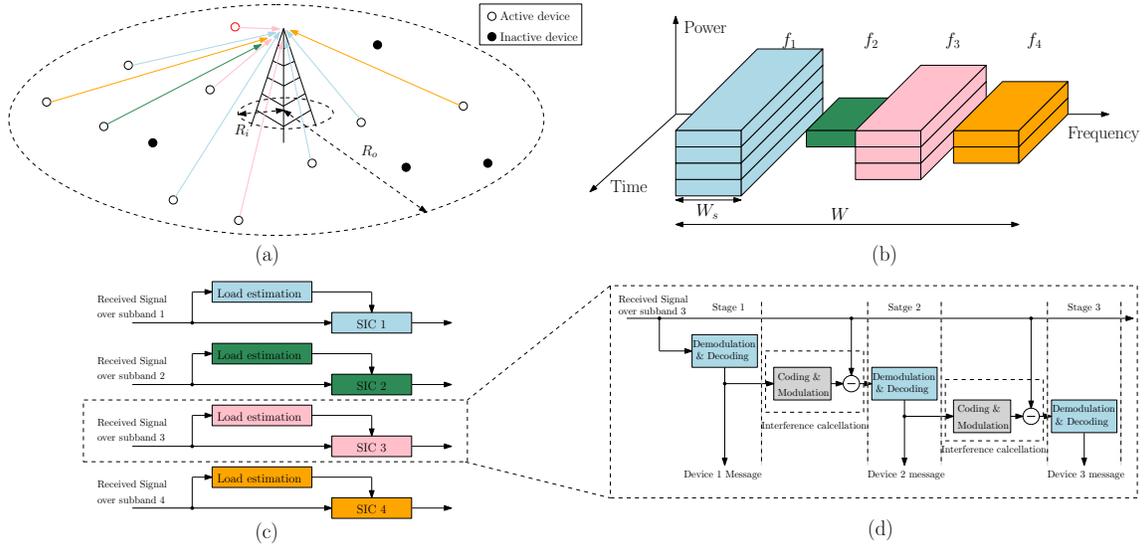}
\caption{Random non-orthogonal multiple access for cellular M2M communications. (a) Each MTC device randomly chooses a subband for its data transmission. (b) The received signals at the base station from different subbands. Devices perform power control so that the received power from all the devices at the base station is the same. (c) The base station performs load estimation and successive interference cancellation over all the subbands. (d) The multi-stage structure of SIC for subband 3, where 3 devices have transmitted their messages over it. }
\label{fig:M2MNOMAfig}
\end{figure*}
\subsection{The Basic Concept of Uplink NOMA}
For simplicity, we assume here two users and a single transmitter and receiver antenna. User $i$ transmit signal $s_i$ with transmission power $p_i$ (see Fig. \ref{fig:NOMA_OMA}-a). The received signal at the base station is represented by:
\begin{align}
y=\sqrt{p_1}h_1s_1+\sqrt{p_2}h_2s_2+w,
\end{align}
where $w$ denotes the received noise including inter-cell interference and $h_i$ is the channel coefficient between user $i$ and the base station. In NOMA, both $s_1$ and $s_2$ are sent over the same frequency band in the same time slot and therefore interfere with each other. At the base station, successive interference cancellation (SIC) is implemented, where first $s_1$ is decoded by treating $s_2$ as interference. Once the receiver correctly decodes $s_1$, it subtracts $s_1$ from the received signal $y$ and then decodes $s_2$. The receiver decides the order of decoding according to the effective SINR of the users.

From an information-theoretic point of view, NOMA with SIC is an optimal multiple access scheme in terms of the achievable multiuser capacity region in both uplink and downlink \cite{higuchi_NOMA}. In NOMA, the performance gain compared to OMA increases when the difference in channel gains or path loss between the users and the base station is large. Fig. \ref{fig:NOMA_OMA}-b shows the capacity region of a single cell scenario with two users with different SNRs, where the total bandwidth is assumed to be 1 Hz \cite{higuchi_NOMA}. Point A for both cases is achieved when the signal for user 1 is decoded first. Point B is achieved when the base station first decodes user 2. It is important to note that NOMA always achieves the highest sum rate while OMA has a gap to the maximum achievable sum rate, with the exception of one point, the symmetric case, where the users achieve the same throughput. It is also clear that when the difference between channel gains increases, NOMA brings greater improvements compared with OMA.

The main benefit of NOMA for IoT is the removal of the need for an RA stage and enabling the devices to transmit in the same channels. This leads to the efficient use of available radio resources, and solves the signaling overhead problem due to conventional RA strategy in cellular systems.
\subsection{NOMA For Massive Cellular IoT}
In the proposed random NOMA for M2M communications, the devices do not need to perform the RA procedure to access the network. Instead, the random access and the data transmission is combined and the devices transmit their messages over randomly selected subbands. This is necessary to minimize the overhead which is critical for many M2M applications with small message sizes. We assume that the total bandwidth is divided into several subbands, and each device which has data to transmit will randomly choose a subband for its data transmission. Multiple subbands have been considered in this paper to show how compatible the proposed NOMA is with the existing LTE and LTE-Advanced standards, which are based on OFDMA technologies. The devices will then choose a channel code with an appropriate code rate and encode their messages along with their terminal identities. The coded message is then sent over the selected subband by the MTC device. Upon receiving the message, the base station performs successive interference cancellation to decode the message of each MTC device over each subband. Fig. \ref{fig:M2MNOMAfig} shows the random NOMA for massive cellular IoT, where only 4 subbands are available.

As the devices randomly choose a subband for their transmissions, the number of MTC devices transmitting over different subbands is not the same. Accordingly the channel code rate cannot be fixed but instead be adapted to the random activities of devices. For this aim, Raptor codes \cite{Raptor}, which are rateless and can generate as many coded symbols as required by the BS, can be used. The code structure is random and can be represented by a bipartite graph; then the BS can reproduce the same bipartite graph using a pseudo random generator with the same seed. When more than one device selects the same seed and transmits over the same subband, they will be transmitting using exactly the same code structure; thus the BS cannot differentiate between them as there is no structural difference between the received codewords. We call this event as \emph{collision}. But we note that a collision requires both the same subband and same code structure to be chosen making it far less likely to occur than a traditional RA collision.
\begin{figure}[t]
\centering
\includegraphics[width=\columnwidth]{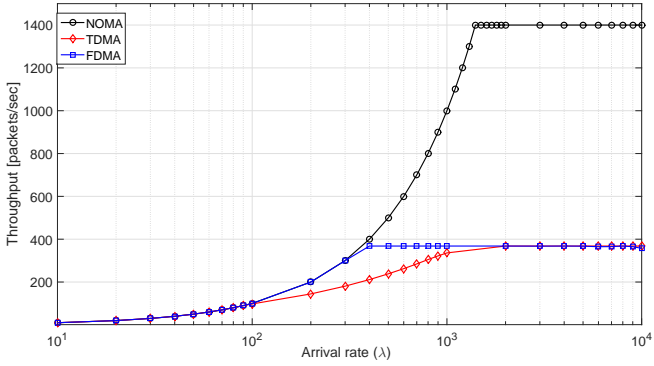}
{\footnotesize (a) NOMA vs. TDMA and FDMA when only one subband is used for NOMA. Minimum subband bandwidth in FDMA is 100kHz and minimum time slot duration in TDMA is 1ms. The total bandwidth is 1 MHz. }
\includegraphics[width=\columnwidth]{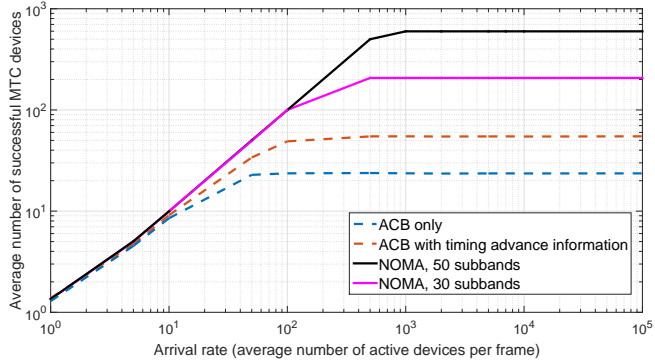}
{\footnotesize (b) Average number of successful MTC devices versus arrival rate. The message length of MTC devices is 1024 bits, the number of preambles is 64, and a resource block has 1 MHz bandwidth and time duration of 1 ms.}
\caption{Throughput of NOMA vs. OMA.}
\label{fig:Delay}
\end{figure}

In Fig. \ref{fig:Delay} we have shown a simple simulation result to show the advantage of NOMA over conventional OMA strategies. As can be seen in Fig. \ref{fig:Delay}-a, NOMA outperforms uncoordinated TDMA and FDMA  strategies and can support a larger number of devices, while FDMA and TDMA suffer from high probability of collision. We compare the proposed random NOMA scheme with different number of subbands with the access class barring (ACB) scheme \cite{ACB_Time} when the same number of radio resources are available in Fig. \ref{fig:Delay}-b. As can be seen in this figure, NOMA can support a significantly larger number of devices compared to ACB schemes. It is important to note that in the ACB scheme, we assume that a device can successfully deliver its message to the base station in its corresponding data channel, when it has successfully completed the RA phase; thus, the only limiting factor for the ACB scheme is the preamble collision in the RA procedure.

\subsection{Potentials of NOMA for Massive Cellular IoT}
NOMA can bring many benefits to cellular systems which include but not limited to the following:
\begin{itemize}
\item Effective use of spectrum and higher system throughput through exploiting the power domain and utilizing non-orthogonal multiplexing.
\item Robust performance gain in high mobility scenarios, where orthogonal multiple access schemes obtains no frequency-domain scheduling gain as channel state information is outdated, but NOMA provide gains in high mobility scenarios as it relies on the channel state information at the receiver side \cite{higuchi_NOMA}.
\item NOMA is compatible with OFDMA and its variants and can be applied on top of OFDMA for downlink and SC-FDMA for uplink \cite{China_NOMA}.
\item NOMA can be combined with beamforming and multi-antenna technologies to improve the system performance \cite{higuchi_NOMA}.
\item NOMA can be easily combined with radio resource management and random access techniques to solve the collision and overload problem in M2M communications.
\item Using clustering and group-based scheduling, NOMA can be used in M2M communications as the multiple access technique to deliver messages of a group of devices to the base station or the cluster head.
\item Using NOMA, the RA procedure can be eliminated and therefore the access delay and signaling overhead will be significantly reduced.
\end{itemize}

\section{Practical Considerations of Massive NOMA for Massive Cellular IoT and Future Directions}
Although NOMA can improve spectrum efficiency and system capacity, there are many practical challenges for this technology to be potentially  used in real wireless systems for M2M communications. Here, we outline the main practical consideration of massive NOMA for M2M communications.

\textbf{\emph{Traffic and load estimation at the base station:}} As the devices randomly select the subbands for their data transmission, the base station needs to accurately estimate the number of devices which are transmitting over each subband. One approachh is to perform power control at the devices so the received power over each subband will be proportional to the number of devices transmitting over that subband. We discuss this in more details in the following.

\textbf{\emph{Channel estimation and power allocation:}} As many devices want to simultaneously communicate with the base station, it is almost impossible for the base station to estimate the channel to all of these devices. The problem becomes more challenging when multiple antennas are used in either the devices or the base station. One could consider the channel between each device and the base station as reciprocal in each direction (this is the case in time division duplexing), so the devices can estimate their channel to the base station using the pilot signal periodically broadcasted by the base station and then adjust their transmission power so the received signal power at the base station is the same fixed value for each device. This is beneficial for M2M communications as the devices have small data packets and the capacity gains of NOMA are not the main advantages of this strategy. Instead, NOMA improves the system throughput through eliminating the RA procedure and enable multiuser detection at the base station.

\textbf{\emph{Synchronization among devices:}} In random NOMA, the devices will be identified during the data transmission, so it is not possible for the base station to determine the timing advance information for every device. One could consider that the devices will estimate their timing information from previous transmissions or according to their location information, which is more practical in M2M applications with fixed location devices. Providing time synchronicity between a large number of devices is a challenging problem and requires major technical efforts.

\textbf{\emph{Proper channel code design:}} The effective data rate for each device is determined by the number of devices which are transmitting in the same subband. As the devices randomly select a subband for the data transmission, the number of devices which are transmitting over each subband is not known beforehand. This means that the code rate at which each device should transmit its data is not fixed. One approach is to use rateless codes, so the devices will transmit using a rateless code and stop their transmission once they received an acknowledgment from the base station. This has been investigated in \cite{Mahyar_SPM_Raptor}, but one should take into account the random structure of rateless codes and think of a way to exchange the random graph structure between the base station and the devices.

\textbf{\emph{Complexity of SIC:}} In existing orthogonal approaches, the BS needs to perform a separate decoding for each device but through the proposed NOMA, the same decoder can be used for all the devices in a sequential manner. However, we can also consider a separate decoder for each device at the decoder and decode each device considering the signal from all other devices as additive noise. This may slightly reduce the throughput but the degradation can be neglected as the achievable rate is mainly determined by the device with the lowest SINR at the BS. interested readers are referred to \cite{Mahyar_SPM_Raptor} for further details on the SIC process.

\textbf{\emph{User fairness:}} The BS can allocate higher bandwidth for those devices which have low quality link to the BS so they can transmit with a lower power. This way the devices with low quality links will transmit with lower power over larger bandwidth and can achieve the same throughput or energy efficiency as the devices with high quality links which are transmitting over a smaller bandwidth or with higher power.

\section{Conclusions}
This article reviewed recent advancement in random access techniques for M2M communications and presented an overview of their benefits and challenges. We have described the basic concept of uplink non-orthogonal multiple access and proposed it as the potential multiple access technology for future cellular systems to accommodate the tremendous growth of M2M applications and traffic. The practical challenges of massive NOMA for M2M communications were also presented and future research directions were highlighted. Massive NOMA offers high throughput efficiency with simple system structure, which are particularly beneficial for massive IoT applications with low-cost low-power and low-complexity devices, and can provide system scalability to support the massive number of devices involved in M2M communications. This technology can be easily adopted by 3GPP technologies for M2M communications to further boost the system performance of current cellar solutions for IoT.

\bibliographystyle{IEEEtran}
\footnotesize
\bibliography{IEEEabrv,References}

\begin{IEEEbiographynophoto}{Mahyar Shirvanimoghaddam} [M] (mahyar.shirvanimoghaddam@newcastle.edu.au) received the B. Sc. degree with 1’st Class Honours from University of Tehran, Iran, in September 2008, the M. Sc. Degree with 1’st Class Honours from Sharif University of Technology, Iran, in October 2010, and the Ph.D. degree from The University of Sydney, Australia, in January 2015, all in Electrical Engineering. He then held a research assistant position at the Centre of Excellence in Telecommunications, School of Electrical and Information Engineering, The University of Sydney, before coming to the University of Newcastle, Australia, where he is now a Postdoctoral Research Associate at the School of Electrical Engineering and Computer Science. His general research interests include channel coding techniques, cooperative communications, compressed sensing, machine-to-machine communications, and wireless sensor networks.
\end{IEEEbiographynophoto}

\begin{IEEEbiographynophoto}{Mischa Dohler} [F] (mischa.dohler@kcl.ac.uk) is a full professor in wireless communications at King’s College London, head of CTR, co-founder and member of the
Board of Directors of the smart city pioneer Worldsensing, Distinguished Lecturer of the IEEE, and Editor-in-Chief of Transactions on Emerging Telecommunications Technologies. He is a frequent keynote, panel, and tutorial speaker. He has pioneered several research fields, contributed to numerous wireless broadband and IoT/M2M standards, holds a dozen patents, organized and
chaired numerous conferences, has more than 200 publications, and authored several books. He has a citation h-index of 37. He acts as a policy, technology, and entrepreneurship adviser, examples being Richard Branson’s Carbon War Room, the House of Lords of the United Kingdom, the EPSRC ICT Strategy Advisory Team, the European Commission, the ISO Smart City working
group, and various startups. He is also an entrepreneur, angel investor, passionate pianist, and fluent in six languages. He has talked at TEDx. He has had coverage by national and international TV and radio, and his contributions have been featured on BBC News and in the Wall Street Journal.
\end{IEEEbiographynophoto}

\begin{IEEEbiographynophoto}{Sarah J. Johnson} [M] (sarah.johnson@newcastle.edu.au) received the B.E. degree in electrical engineering, and the Ph.D. degree from the University of Newcastle, Australia, in  2000 and 2004, respectively. She held a postdoctoral position at NICTA, Australia’s Information and Communications Technology Research Centre of Excellence, before returning to the University of Newcastle, where she is now an Associate Professor and Australian Research Council Future Fellow.  Her research interests include error correction codes and information theory for multiple-user communication networks. She has authored a book entitled Iterative Error Correction (Cambridge University Press).
\end{IEEEbiographynophoto}

\end{document}